\documentclass{emulateapj}
\submitted{{\it Submitted for publication in ApJL}}
\usepackage {graphicx}
\usepackage{hyperref}
\usepackage{amsmath} 
\usepackage{amssymb} 
\usepackage{graphics}
\usepackage{epsfig}  
\usepackage{float}
\def\be{\begin{equation}}
\def\ee{\end{equation}}
\def\ba{\begin{eqnarray}}
\def\ea{\end{eqnarray}}

\newcommand{\hMpc}{{\ifmmode{h^{-1}{\rm Mpc}}\else{$h^{-1}$Mpc }\fi}}  
\newcommand{\hGpc}{{\ifmmode{h^{-1}{\rm Gpc}}\else{$h^{-1}$Gpc }\fi}}  
\newcommand{\hmpc}{{\ifmmode{h^{-1}{\rm Mpc}}\else{$h^{-1}$Mpc }\fi}}  
\newcommand{\hkpc}{{\ifmmode{h^{-1}{\rm kpc}}\else{$h^{-1}$kpc }\fi}}  
\newcommand{\hMsun}{{\ifmmode{h^{-1}{\rm {M_{\odot}}}}\else{$h^{-1}{\rm{M_{\odot}}}$}\fi}}  
\newcommand{\hmsun}{{\ifmmode{h^{-1}{\rm {M_{\odot}}}}\else{$h^{-1}{\rm{M_{\odot}}}$}\fi}}  
\newcommand{\Msun}{{\ifmmode{{\rm {M_{\odot}}}}\else{${\rm{M_{\odot}}}$}\fi}}  
\newcommand{\msun}{{\ifmmode{{\rm {M_{\odot}}}}\else{${\rm{M_{\odot}}}$}\fi}}  
\newcommand{\kms}{{\ifmmode{{\mathrm{\,km\ s}^{-1}}}\else{\,km~s$^{-1}$}\fi}}

\newcommand{\bulla}{1E0657---56} 
\newcommand{\bullg}{SL2S J08544-0121}
\shorttitle{Bullet groups}
\shortauthors{Fern\'andez-Trincado et al.}

\begin{document} 

\title{The abundance of Bullet-Groups in $\Lambda$CDM}
\author{J. G. Fern\'andez-Trincado$^{1,2,3}$, J. E. Forero-Romero$^1$,
  G.Foex$^4$, T. Verdugo$^3$ and V. Motta$^4$} 
\affil{$^1$ Departamento de F\'{i}sica, Universidad de los Andes,
  Cra. 1 No. 18A-10, Edificio Ip, Bogot\'a, Colombia\\ 
       $^2$ Institute Utinam, CNRS UMR6213, Universit\'e de
  Franche-Comt\'e, OSU THETA de Franche-Comt\'e-Bourgogne,
  Besan\c{c}on, France\\ 
       $^3$ Centro de Investigaciones de Astronom\'ia, AP 264,
  M\'erida 5101-A, Venezuela\\
  $^4$ Instituto de F\'isica y Astronom\'ia, Universidad de
  Valpara\'iso, Avda. Gran Breta\~na 1111, Playa Ancha, Valpara\'iso
  2360102, Chile  
}
\email{jfernandez@obs-besancon.fr}
\email{je.forero@uniandes.edu.co}
\begin{abstract}

We estimate the expected distribution of displacements between the two
dominant dark matter (DM) peaks (DM-DM displacements) and between
DM and gaseous baryon peak (DM-gas displacements) in dark matter halos 
with masses larger than $10^{13}$\hMsun. We use as a benchmark the
observation of \bullg, which is the lowest mass system ($1.0\times
10^{14}$\hMsun) observed so far
featuring a bi-modal dark matter distribution with a dislocated
gas component. We find that $(50 \pm 10)\%$ of the dark matter
halos with circular velocities in the range $300\kms$ to $700\kms$
(groups) show DM-DM displacements equal or larger than $186 \pm
30$\hkpc as observed in \bullg. For dark matter halos with circular
velocities larger than $700\kms$ (clusters) this fraction rises to
$(70 \pm 10)\%$. Using the same simulation we estimate the DM-gas
displacements and find that $0.1$ to $1.0\%$ of the groups should
present separations equal or larger than $87\pm 14$\hkpc corresponding
to our observational benchmark; for clusters this fraction rises to
$(7\pm 3)\%$, consistent with previous studies of dark matter to
baryon separations. Considering both constraints on the DM-DM and
DM-gas displacements we find that the number density of groups
similar to \bullg\ is $\sim 6.0\times 10^{-7}$ Mpc$^{-3}$, three times
larger than the estimated value for clusters. These results open up
the possibility for a new statistical test of $\Lambda$CDM by looking
for DM-gas displacements in low mass clusters and groups. 
\end{abstract}


\section{Introduction}

The Bullet Cluster (\bulla) provided a new kind of observational
evidence for the existence of dark matter
\citep{Markevitch2004,Clowe2006}. Since then it has been used to test
the Cold Dark Matter (CDM) paradigm itself by quantifying different
aspects such as the expected displacement between the dominant dark
matter and baryonic component in a $\Lambda$CDM Universe
\citep{ForeroRomero2010}, the substructure velocity required to
produce such displacement
\citep{Milosavljevic2007,Springel2007,Mastropietro2008} and its
abundance in large N-body cosmological simulations
\citep{Hayashi2006,Lee2010, Thompson2012}. It has also been used to
constrain the dark matter particle self-interaction cross section and
to explore possible extensions to the concordance cosmological model
\citep{Farrar2007,Lee2012}.

Since then, other examples of Bullet-like systems have been found; 
MACS J0025.4-1222 \citep{Bradac2008}, Abell 2744 \citep{Merten2011},
DLSCL J0916.2+2951 \citep{Dawson2012}, ZwCl 1234.0+02916
\citep{Dahle2013}. Recently \cite{Gastaldello} observed a DM-gas
displacement of $87\pm 14$\hkpc and a DM-DM separation of $186\pm
30$\hkpc in \bullg, a low mass cluster system with a total
mass $2.4\pm 0.6 \times 10^{14}\Msun$ found in the the Strong Lensing
Legacy Survey (SL2S) sample \citep{Cabanac2007,More2012}.

Using a Sheth-Mo-Tormen mass function at $z=0$, one can estimate that
systems around this mass are $\sim 65$ times more abundant than massive
clusters in the mass range of the Bullet Cluster $>10^{15}$\hMsun
\citep{Sheth2001,hmfcalc}. This  should open up the possibility of
finding Bullet-like groups in large numbers to test
$\Lambda$CDM. However, a larger abundance of small mass systems has to
be weighted by the probability of being in a merger and  presenting a
large displacement between the DM and gaseous components.  These two
conditions (merger rates, maximum possible displacement) are a function
of DM halo mass in $\Lambda$CDM cosmologies.  A detailed statistical
study to estimate the DM-gas displacements of bullets has been
performed for clusters \citep{ForeroRomero2010} but not for lower mass
systems. 

In this Letter we extend such study for systems in the group mass
scale. We measure the abundance of DM systems with a multi-modal
morphology (large DM-DM displacements) and estimate the amount of
systems with a Bullet-like configuration  (large DM-gas
displacements). To this end we use a high resolution N-body
cosmological simulation (Bolshoi) that allows us to find multi-modal
dark matter distributions in hosts with circular velocities larger
than $300$\kms ($\sim 1.0\times 10^{13}\hMsun$).

This Letter is organized as follows. In Section
\ref{sec:simulation} we present the simulation and the halo
catalogs. We continue in Section \ref{sec:setup} with the  geometry of
the problem at hand and the measurement setup. Next in   Section
\ref{sec:results} we present our results and observational
perspectives to finally conclude in Section \ref{sec:conclusions}.  

\section{Simulation, halo catalogs and pairs}
\label{sec:simulation}

We use the Bolshoi Run, a cosmological DM only simulation over a cubic
volume of 250\hMpc comoving on a side \citep{2011ApJ...740..102K}. The
simulation uses the ART code \citep{Kravtsov1997} to follow the
evolution of a dark matter density field from $z=80$ to $z=0$ sampled
with $2048^3$  particles. The cosmology used  corresponds to  the
spatially flat concordance model with the following parameters:  the
density parameter for matter (dark matter and baryons)
$\Omega_m=0.27$, the density parameter for baryonic matter
$\Omega_b=0.0469$, the density parameter for dark energy
$\Omega_{\Lambda}=0.73$, the Hubble parameter $h=0.7$, the slope of
the primordial power spectrum $n=0.95$ and the amplitude of mass
density fluctuations (at redshift z$=$0) $\sigma_8=0.82$.  These
cosmological parameters are consistent with the nine-year Wilikinson
Microwave Anisotropy Probe (WMAP) results \citep{hinshaw_etal13}. A
detailed presentation of the simulation can be found in
\citet{2011ApJ...740..102K}.  

This results in a mass resolution of $1.35 \times 10^8$
M$_{\odot}$h$^{-1}$ for each computational particle. The completeness
limit in this simulation is set for halos with $100$ particles
corresponding to a mass of $1.35\times10^{10}$\hMsun\ or a maximum
circular velocity $V_{c}$ of $50$\kms. 

We use DM halo catalogs constructed using the Bound Density Maxima (BDM)
algorithm \citep{BDM,BDMb}. To define the radius of a halo we use a
density threshold of 360 times the mean density of the Universe. An
important feature of BDM is that it allows us to detect sub-halos
inside larger virialized structures.     

All the raw data used in this Letter are available through the
Multidark database \footnote{\url{www.multidark.org}}
\citep{Riebe2013}.  Furthermore, in order to facilitate the
reproducibility and reuse of our results we have made available all
the data and the source code available in a public
repository \footnote{\url{https://github.com/Fernandez-Trincado/Bullet\_Groups-2014}}.  
 
To construct our main halo sample we follow three steps. Firstly, we
select all the host halos (i.e. halos that are not inside a larger
halo) with circular velocities $V_{\rm c}\geq 300$\kms ($\geq 1\times
10^{13} \hMsun$). Secondly, we select the sub-halos with circular
velocities $V_{\rm c}\geq 75$\kms ($\geq 5\times
10^{10}\hMsun$). Thirdly, we associate each host halo to its most
massive sub-halo.  

Each one of the pairs (host \& sub-halo) is considered as a
potential Bullet-like system and is kept for the analysis described in the
next Section  by using the sub-halos as the tracer of the sub-dominant
dark matter clump in the merging cluster, i.e. the bullets.  

Finally, we split the sample into two  populations; groups, with
$300\kms < V_{\rm c, host}<700\kms$; and clusters, with $V_{\rm
  c,host}>700\kms$ ($\geq 8.0\times 10^{13}$\hMsun). These two samples
are constructed at four redshifts: $z=0.0, 0.25, 0.5$ and $1.0$.  The
number of pairs at each redshift in each population is listed on the
first row of Table \ref{table:numbers}.

\section{Bullet Geometry and Measurement Setup}
\label{sec:setup}

Bullet-like configurations have two dominant dark matter structures:
the host halo and the dominant sub-halo. We describe the kinematics of
this configuration by the  position and velocity vectors of the
sub-halo in a frame of reference where the main halo is at rest; thus
$\vec{v{}}=\vec{v}_{\rm sub}-\vec{v}_{\rm host}$ and
$\vec{r{}}=\vec{r}_{\rm sub}-\vec{r}_{\rm host}$, where the subscripts $host$
and $sub$ refer to the host and sub-halo, respectively. The positions
$\vec{r}$ correspond to the minimum of potential and the velocities
$\vec{v}$ are the center of mass velocities.

The angle between these two vectors can be quantified by, 
\begin{equation}
  \mu\equiv
  \cos(\theta)=\frac{\vec{v{}}\cdotp{}\vec{r}}{\left\|\vec{v}{}\right\|
    \left\|\vec{r}\right\|} .
 \end{equation} 
This encodes information about the collision, i.e. cases of
$|\mu|\sim 1$ can be considered as head-on collisions while
$|\mu|< 0.9$ describes a grazing trajectory.  

The geometrical configuration can be further described by the
the circular velocity of each component and the
size of the host halo $R_{\rm vir}$. Another useful quantity computed
in the simulation is the distance between the minimum of potential for
the host halo and its center of mass (computed from all the particles
inside the $R_{\rm vir}$), $X_{\rm
  off}=||\vec{r}_{min}-\vec{r}_{cm}||/R_{\rm vir}$, which serves as a
measurement of how much the host halo is perturbed.  

In this Letter we work with two quantities that could be inferred
from observations of Bullet-like systems. The projected distance
between two dominant DM clumps, $d_{\rm 2D}$, and the projected distance
between the DM and the gas clumps, $d_{\rm 2D}^{\rm bar}$. The
projection is computed along the $z$-axis for all halos at all
redshifts.

From the simulation point of view, the first quantity can be
translated into the 2D projected values of $||\vec{r}||$ and its value
relative to the virial radius $D_{\rm   off}= ||\vec{r}||_{\rm
  2D}/R_{\rm vir}$. The second quantity, the projected DM-gas
distance, is not directly available from a DM-only simulation but can
be estimated from the data.  

We also use the physical quantities described above to discriminate
three main stages in a Bullet-like encounter with $|\mu|\sim
1$. First, when the sub-halo crosses the virial radius of the host
halo starting a head on collision, $D_{\rm   off} \sim 1$ and $\mu\sim
-1$. Second, as the sub-halo crosses for the first time the center of
the host halo $D_{\rm off}<1.0$ and $\mu\sim 1$. Third, as the
sub-halo reaches apogee and comes back to the center of the halo
$D_{\rm off} < 1.0$ and $\mu\sim -1$.

\section{Results}
\label{sec:results}

\subsection{DM-DM Displacements}
\label{fig:displacement}

\begin{figure*}
\begin{center}
\includegraphics[width=0.9\textwidth]{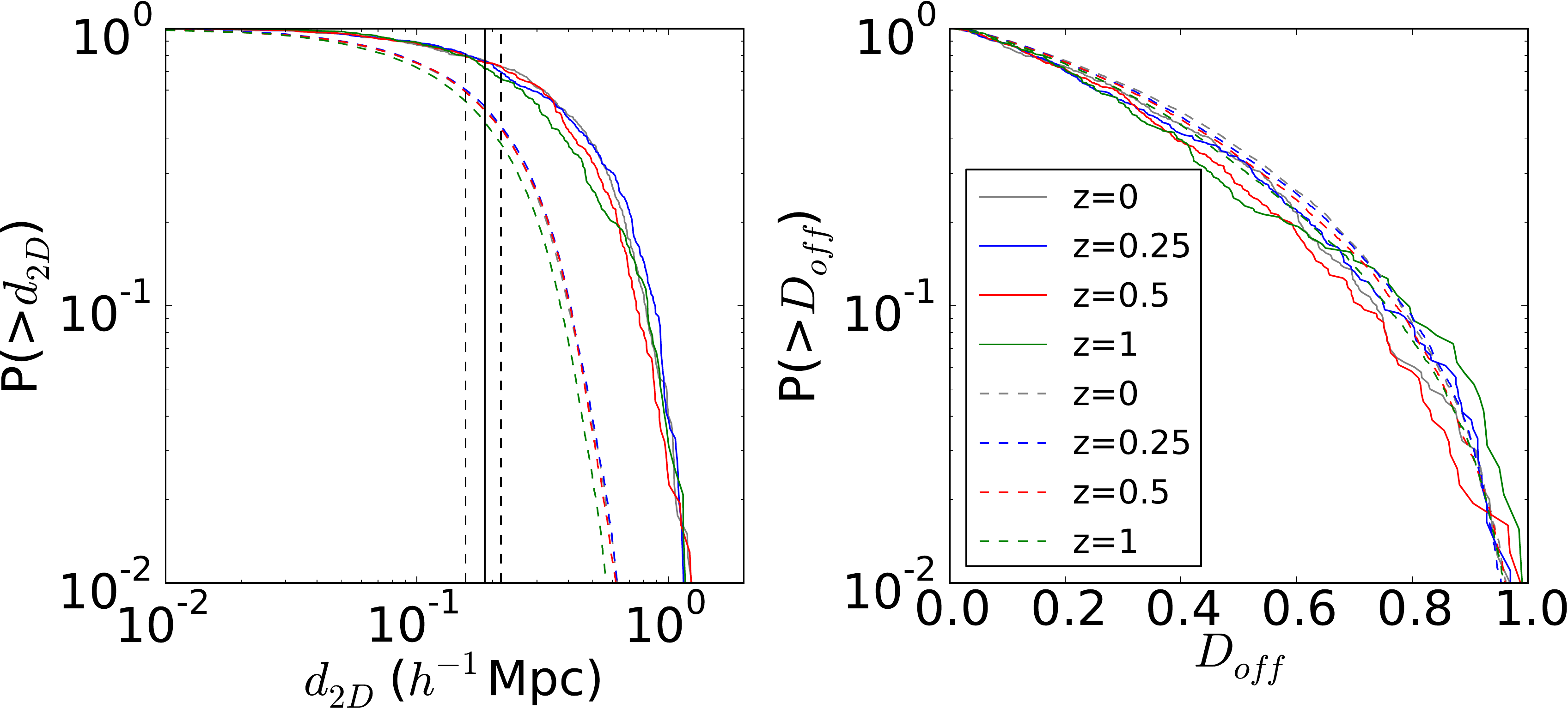}
\end{center}
\caption{
  Integrated probability distribution for the displacement between the
  center of the host halo and its dominant sub-halo at all redshifts
  $z=0$, $0.25$, $0.5$ and $1.0$. The left panel
  shows the results in terms of the physical displacements while the
  right panel shows the displacements normalized by the virial radius
  of the host halo. The continuous (dashed) line corresponds to the halos in the
  cluster (group) sample.
  The vertical lines show the mean value and uncertainties ($186\pm
  30\hkpc$) in the  separation between the two dark matter clumps
  estimated in \citet{Gastaldello} for the \bullg. Between $40\%$
  to $60\%$ of the groups show a displacement equal or larger than
  this observational benchmark. This fraction rises to $60\%$ and
  $80\%$ in clusters.}
\label{fig:displacement}
\end{figure*}

Figure \ref{fig:displacement} presents the integrated
probability distribution for the DM-DM displacements, $d_{\rm 2D}$. The
left panel shows the displacement in physical units and the right
panel as a fraction of the virial radius of the host halo.  The panel
with the projected 2D physical displacements also shows a vertical
stripe with the estimated displacement for the bullet group reported
by \cite{Gastaldello}.   

In the group sample we see that a fraction of
$40\%$ to $60\%$  should present a displacement equal than the
estimate for \bullg; in the cluster sample this fraction increases to
$70\%$-$80\%$. The uncertainties in this estimates are derived from
the uncertainties in the displacement measurement for \bullg. This
fraction is naturally higher in more massive systems because they are
larger in size. Normalizing the displacements by the virial radius,
right panel Figure \ref{fig:displacement}, we see that the
distributions are similar for the two samples at all redshifts.

\subsection{Collision Geometries}
\label{sec:geometry}

\begin{figure*}
\begin{center}
\includegraphics[width=1.0\textwidth]{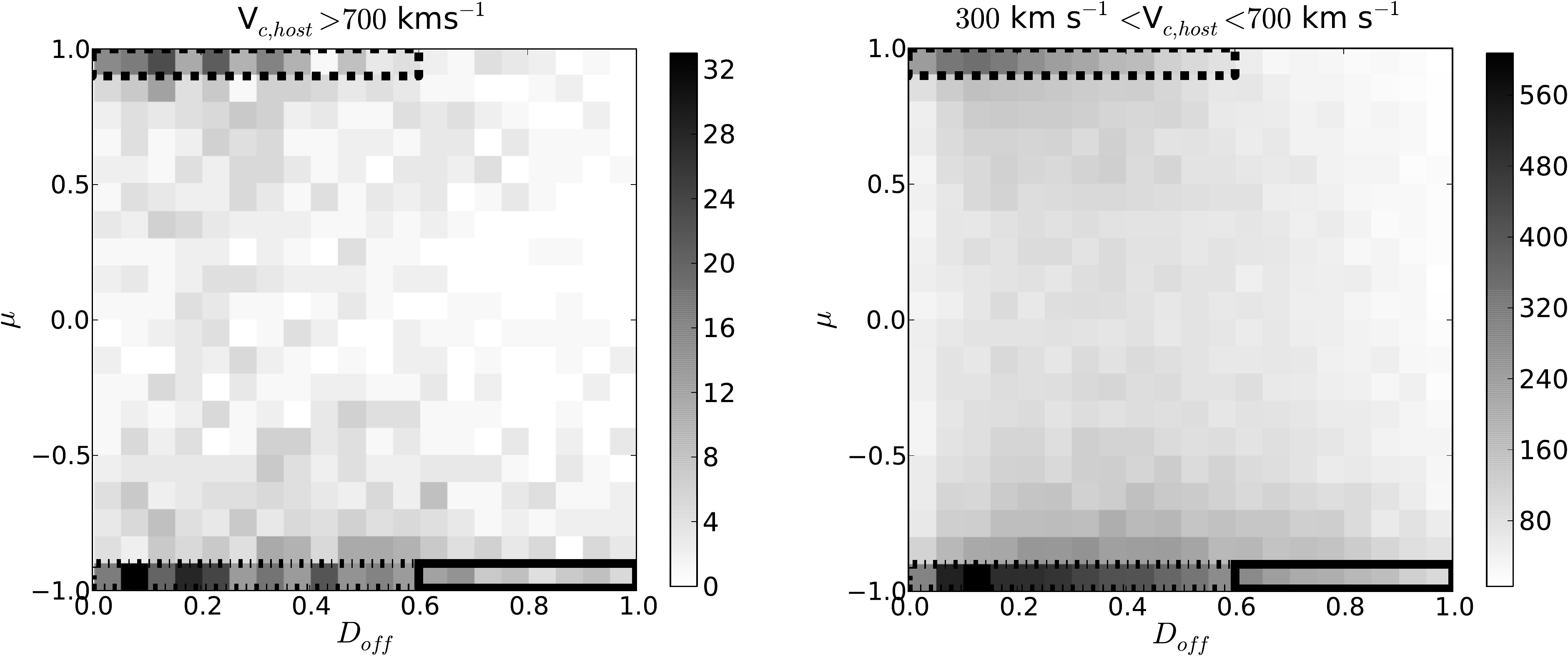}
\end{center}
\caption{2D histograms in the plane $\mu$-$D_{\rm off}$ describing the
  geometry of the bullet encounter. The left panel corresponds to
  clusters and the right panel to groups.  The weak redshift evolution
  of the structure in these planes allows us to include halos at all
  redshifts to construct these figures. The three rectangles
  (continuous, dashed, dot-dashed) roughly delimit stages of interest
  in a head-on collision (initial infall, first crossing until apogee,
  return after apogee) as described in Section 
  \ref{sec:geometry}.}  
\label{fig:geometry}
\end{figure*}

Figure \ref{fig:geometry} presents the geometry of the interactions using the 
variables $\mu$ and $D_{\rm off}$. The first evident feature is that
most of the configurations have head-on encounters, $|\mu|>0.9$
($\theta\leq 30^{\circ}$), while only a minority with $|\mu|<0.9$ can
be described as having  grazing trajectories. 

For pairs on radial trajectories there are three regions of interest
in this plane that correspond to the three merging stages described
at the end of Section \ref{sec:geometry} assuming that the sub-halo
merges (or falls below the BDM detection threshold) right at its
second pass through the center of the host halo \citep{Poole2006}.
These regions are shown as three different rectangles in Figure
\ref{fig:geometry}. 

The first region (continuous rectangle) has $\mu<-0.9$ and $D_{\rm
  off}>0.6$, which locates the systems where a head-on collision has
just started. The second region (dashed rectangle) has $\mu>0.9$ and
$D_{\rm off}<0.6$; at this stage the collision continues after the
first crossing of the host's center. The low number of halos with
radial infalling velocities and displacements $D_{\rm off}>0.6$
suggests that this the maximum range of radii for the apogee.  The
third region (dot-dashed rectangle) corresponds to $\mu<-0.9$ and
$D_{\rm off}<0.6$ that is the the secondary infall after apogee. 

In the next subsection we use the information in this collision
sequence to estimate the expected DM-gas displacement. 

\subsection{DM-gas Displacements}
\label{sec:baryonic_displacements}

The results we have derived so far apply to multi-modal systems and
their expected separation between the two dominant dark matter
clumps. However, a non-zero DM-DM displacement does not imply 
a non-zero DM-gas displacement. We now estimate these displacements
from the available information in our DM only simulation.

Our estimates are based on the different kinds of trajectories and
collision stages described in previous sections. To start with, we
consider that systems with $|\mu|<0.9$ describing grazing
trajectories have a DM-gas displacement equal to zero.  In systems
with head-on collisions, $|\mu|>0.9$ the systems with $\mu<-0.9$ and
large displacements $D_{\rm   off}>0.6$ most probably describe the
beginning of the interaction and should have DM-gas displacements
equal to zero as well. In all other cases the DM-gas displacement
should be different from zero with the approximated value
$d_{\rm   2D}^{\rm bar} = X_{\rm off} R_{\rm   vir}$, where $X_{\rm   off}$ is
the offset computed between the minimum of potential and the center of
mass for each host halo computed from all the matter inside the virial
radius.

This simplified model does not take into account that there is a
fraction of halos with $\mu<-0.9$ and $D_{\rm off}<0.6$ for which the
collision has not started and should have $d_{2D}^{\rm bar}=0$. A detailed
modeling of this fraction requires the study of the complete merging
history  of the halo and sub-halo, a study beyond the scope of this
Letter. Instead we caution the reader that the derived fraction of
halos with a displacement $<d_{2D}^{\rm bar}$ must be considered as an
upper limit. 

\begin{figure*}
\begin{center}
\includegraphics[width=0.8\textwidth]{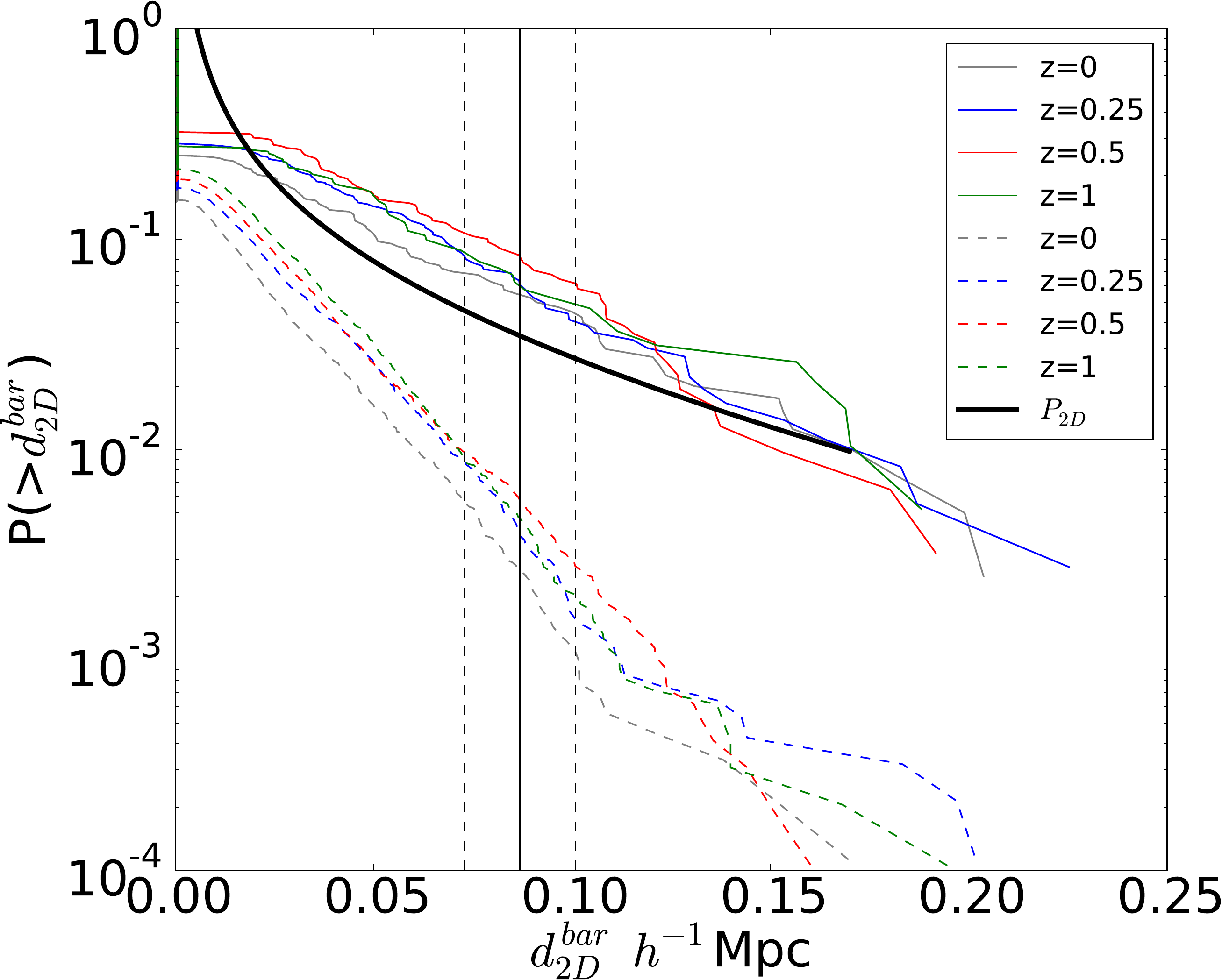}
\end{center}
\caption{Integrated probability distribution for the estimated
  DM-gas displacements in the group and cluster samples. Continuous (dashed)
  lines correspond to clusters (groups). The continuous black line
  marked as $P_{2D}$ shows the statistics reported by
  \citet{ForeroRomero2010} for a cosmological simulation including DM
  and gas. The vertical lines correspond to the mean value and
  uncertainty of the displacement measured for \bullg.} 
\label{fig:baryonic_displacements}
\end{figure*}

The results for the integrated distributions for $d_{\rm 2D}^{\rm bar}$
are shown in Figure \ref{fig:baryonic_displacements}. The dashed lines
represented the results for groups and the continuous lines correspond
to clusters. As a test of our model we compare the cluster results
against the analytic fit provided by \cite{ForeroRomero2010}. This fit
reproduces the statistics for the DM-baryon separation found for
clusters more massive than $>10^{14}$\hMsun\ in a simulation
which included a description for DM and gas with $8$ times more volume
and $8$ times less mass resolution than the Bolshoi Simulation. The
fit is valid for separations larger than $70\hkpc$, beyond which we
find that it provides a remarkably good description within a factor of
$\sim 2$ of our results. This gives us confidence in our approach to
correctly estimate the expected fraction of groups with a DM-gas
displacement. 

Finally, from Figure \ref{fig:baryonic_displacements} we see that only a
fraction of $0.1\%$ to $1\%$ of the groups are expected to have a
DM-gas displacement equal or larger than $87\pm 14$\hkpc as
observed in \bullg. This fraction rises to $4\%$ to $10\%$ in the case
of clusters, consistent with the results reported by
\cite{ForeroRomero2010}.

\subsection{Towards a Statistical Comparison Against Observations}

Recently \cite{Foex2013} presented an analysis of $80$ galaxy groups
in the SL2S sample. From the light distribution, only $34$ objects
($\sim 42\%$) have regular isophotes, $33$ had elongated
isophotes (hints of merging system), and $13$ ($\sim 16\%$) had a
clear bi-modal light distribution; \bullg\ is one of these
$13$ systems.   

The bi-modal objects are defined to have at least a clear second
luminosity peak within $350$\hkpc from the main halo, as traced by the
strong lensing system.  The lowest separation in those systems is
$64$ \hkpc and the average is $145\pm 52$\hkpc. 

We now make a comparison of these fractions against the results of our
simulations. The results are summarized in Table
\ref{table:numbers}. The first row indicates the total number of halos
in each sample at each redshift. The second row shows the number of
objects with DM-DM displacements $46\hkpc< d_{\rm 2D}< 350$\hkpc. The
last row indicates the number of objects in the previous sub-sample
with DM-gas displacements $d_{\rm 2D}^{\rm bar}>87$\hkpc. 

Considering that the statistics for the cluster sample are dominated
by objects in the mass range of the \bullg\ we make a comparison
against this sample. Table \ref{table:numbers} shows that $\sim 40\%$ of the 
clusters are expected to have large DM-DM displacements, which is a
factor of $\sim 2$ larger than the observational estimate by
\cite{Foex2013}. However, roughly $\sim 7\%$ of these systems present a
displacement equal or larger than the observed in the \bullg, a
fraction that is compatible with the $1/13\sim 0.07$ fraction in the
SL2S sample from which \bullg\ was drawn. 

This rough comparison  shows that our estimates for the relative number of
bullet systems (large DM-gas displacements) with respect to
multi-modal systems (large DM-DM displacements) is compatible with
observations. A proper comparison must take into account all the observational
uncertainties, biases and mixes between our two populations (groups \&
clusters at different redshifts) to derive a stronger bound from
the simulation, something that is beyond the scope of this Letter.   

Nevertheless, comparing the absolute number of systems with
characteristics similar to \bullg\ in the group sample
(last row of Table \ref{table:numbers}) we predict that its number
density is $\sim 6.0\times 10^{-7}$ Mpc$^{-3}$, three times larger
than the expected number density of \bullg\ systems in the cluster
sample.

\begin{table*}
\begin{center}
\begin{tabular}{ccccccccc}\hline\hline
        & Groups & Groups & Groups & Groups & Clusters & Clusters & Clusters & Clusters\\
  & $z=0$   & $z=0.25$   & $z=0.5$   & $z=1.0$   & $z=0$   & $z=0.25$   & $z=0.5$   & $z=1.0$ \\
  & ($\# $)    & ($\# $)    & ($\# $)    & ($\# $)    & ($\# $)    & ($\# $)    & ($\# $)    & ($\# $)  \\\hline
Full Sample & 9641 & 9984 & 10244 & 10190 & 400  & 363 & 310 & 192 \\
$64 <d_{\rm 2D}/\hkpc< 350$ & 6188 & 6422 & 6635 & 6933 & 151 &
141 & 120 & 99\\
($64 <d_{\rm 2D}/\hkpc< 350 $) \& ($d_{\rm 2D}^{\rm bar}/\hkpc >87$) & 14 & 25 & 44 & 35 & 8 & 9 & 13 & 8 \\\hline\hline
\end{tabular}
\caption{Absolute number of objects in the groups and clusters sample
  at all redshifts for different selection criteria. }
\label{table:numbers}
\end{center}
\end{table*}

\section{Conclusions}
\label{sec:conclusions}

In this Letter we estimated the fraction of galaxy groups and clusters
in a $\Lambda$CDM cosmology that could present observational features
associated to a Bullet-like event. This is motivated by the recent
observational results of \cite{Gastaldello} where a system (\bullg)
on the mass range $1\times 10^{14}$\hMsun\ and velocity dispersion
$650$\kms was reported to feature a displacement between its baryonic
(gas) and dark matter components. 

We computed the distribution of projected displacements
between the dominant DM clumps in two kinds of systems; groups with
circular velocities $300\kms<V_{\rm c}<700\kms$ ($1.0\times
10^{13}\hMsun < M_{\rm vir}<8.0\times 10^{13}\hMsun$) and clusters with $V_{\rm
  c}>700$\kms ($M_{\rm vir}>8.0\times 10^{13}\hMsun$). We reported these
results at four different redshifts $z=0.0,0.25,0.5$ and $1$. Our
results are based on large DM-only N-body cosmological simulation with
a resolution that allows us to study for the first time Bullet-like
configurations in the mass range of galaxy groups. 

Our main result is that a fraction of $40\%$-$60\%$ of the halos
in the group sample presents DM-DM displacement equal or larger than the
observed displacement for \bullg. For halos in the cluster sample this
fraction increases to $60\%$-$80\%$. We also derived an
estimate for the DM-baryon displacement. In the group sample
$0.1\%$-$1.0\%$ of the halos show a displacement equal or larger than
the  measurements of \bullg\ by \cite{Gastaldello}; in the cluster
sample this fraction rises to $4\%$-$10\%$. 

For the case of \bullg\ a fair comparison is achieved against our cluster
sample which has statistics dominated by objects of similar mass. In
a rough comparison using the observational criteria \citep{Foex2013,
  Gastaldello} we find that the relative number of Bullet-like systems
(large DM-gas displacement) with respect to a general sample of
multi-modal systems (large DM-DM displacements) is consistent with
observations; both are in the range $\sim 7\%$. 

Using the same criteria we find that in the simulation there are
$\sim 6.0\times 10^{-7}$ Mpc$^{-3}$ groups similar to \bullg. This
number density is three times larger than the computed value for
clusters. This opens up a new observational possibility with surveys
such as SL2S that target a large number of groups and estimate its
multi-modal nature from lensing analysis \citep{Foex2013}. An approach
that can be further exploited with upcoming lensing surveys (e.g. with
 he {\it Euclid} satellite) and pushes for X-ray surveys with higher
sensibility: chances are larger to find bullets in systems with lower
X-ray luminosities.\\

We thank the referee for a detailed report that improved the clarity of
this Letter. 

The CosmoSim database used in this paper is a service by the
Leibniz-Institute for Astrophysics Potsdam (AIP). The  Bolshoi
simulation was performed within the Bolshoi project of the University
of California High-Performance AstroComputing Center (UC-HIPACC) and
was run at the NASA Ames Research Center. 

J.G.F-T acknowledges support from Universidad de los Andes de Bogot\'a
- Colombia and Centro de Investigaciones de Astronom\'ia (CIDA) -
Venezuela. 

J.E.F-R acknowledges support from Vicerrector\'ia de
Investigaciones through a FAPA starting grant.

G.F. acknowledges support from FONDECYT through grant 3120160 and
ECOS/CONICYT through grant C12U02.  

T.V. acknowledges support from CONACYT through grant 165365 and
203489 through the program Estancias posdoctorales y sab\'aticas al
extranjero para la consolidaci\'on de grupos de investigaci\'on.  

V.M. gratefully acknowledges FONDECYT support through grant 1120741
and ECOS/CONICYT through grant C12U02.

\bibliographystyle{apj}

\end{document}